# Risk analysis of Trojan-horse attacks on practical quantum key distribution systems

Nitin Jain, Birgit Stiller, Imran Khan, Vadim Makarov, Christoph Marquardt, and Gerd Leuchs

*Abstract*—An eavesdropper Eve may probe a quantum key distribution (QKD) system by sending a bright pulse from the quantum channel into the system and analyzing the back-reflected pulses. Such Trojan-horse attacks can breach the security of the QKD system if appropriate safeguards are not installed or if they can be fooled by Eve. We present a risk analysis of such attacks based on extensive spectral measurements, such as transmittance, reflectivity, and detection sensitivity of some critical components used in typical QKD systems. Our results indicate the existence of wavelength regimes where the attacker gains considerable advantage as compared to launching an attack at 1550 nm. We also propose countermeasures to reduce the risk of such attacks.

*Index Terms*—Electrooptic modulators, Fiber optics, Infrared detectors, Infrared spectra, Optical pulses, Optical receivers, Photodetectors, Quantum key distribution, Quantum cryptography, Reflectometry, Supercontinuum, Transmittance, Trojan-horse attack

## I. INTRODUCTION

As of today, quantum key distribution (QKD) is one of the most promising applications of quantum information technology [1-3]. In its most basic form, QKD facilitates two parties Alice and Bob to exchange a key for encrypting their messages in an information-theoretically secure manner. In theory, any eavesdropping during their key exchange 'protocol' introduces errors in the measured data, thus disclosing the presence of their adversary Eve. If the amount of errors is not too high, Alice and Bob can try to distill a shorter but secret key. Otherwise, they abort the QKD protocol and try to use another quantum channel (or communicate later). The confidentiality of their messages is thereby never compromised.

In practice however, the theoretical 'security' model may not be properly implemented, or the security proof, which is at the heart of this security model may be based on incorrect or insufficient assumptions. In such cases, the eavesdropper may perform an attack and acquire (partial or full) knowledge of the key without causing too many errors, thus breaching the security of the QKD system.

Such theory-practice deviations usually arise due to technical deficiencies and operational imperfections in the system hardware or firmware. The field of *quantum hacking* investigates such deviations [4-6] and many proof-of-principle quantum hacking attacks have been devised and performed on practical QKD systems in the last decade [7-21].

A vast majority of current QKD implementations are fiber-optical, i.e., different system components are connected together by fiber-optic interfaces. For ideal components and interfaces, all of the input light may be perfectly transmitted to the output. But in reality, while traveling through an interface or inside a component, a non-zero portion of the light will be reflected or scattered back. To be more precise, a change of refractive index during propagation induces Fresnel reflection, while density fluctuations in the material of the optical fibers cause Rayleigh or Brillouin scattering. The amount of reflection and scattering may depend on the wavelength and intensity of the input light.

A bright pulse launched from the quantum channel (by Eve) into a QKD subsystem, e.g., Alice, would encounter multiple reflection and scattering sites. A stream of reflected pulses can therefore be expected to propagate out of Alice's device on the quantum channel. By carefully analyzing this *back-reflected* light, Eve could acquire knowledge of the properties and functionality of a component inside Alice. Fig. 1(a) shows Eve launching such an attack from the quantum channel with the intention of knowing the bases selected by Alice.

The first hint of such a *large pulse attack* was provided in Ref. [22]. The basic ideas and applicability conditions for implementing a realistic large pulse attack were extensively investigated one year later [23]. This work reported the results of a simple experimental interrogation of the 'static' settings of Alice and Bob devices and also proposed several realistic countermeasures. In 2006, these ideas were generalized under the name of *Trojan-horse attacks* [24]. These attacks were again experimentally investigated, albeit still in a static sense, with a focus on the plug-and-play architecture [22, 25, 26].

Recently, both realistic and real-time Trojan-horse attacks on practical QKD implementations have attracted considerable attention [19-21]. The implementations include commercial QKD systems, such as Clavis2 from ID Quantique [27], and Cygnus from SeQureNet [28]. Notably, the wavelength chosen

This paragraph of the first footnote will contain the date on which you submitted your paper for review. This work was supported in part by the CHIST-ERA project HIPERCOM, Industry Canada, University Graduate Center in Kjeller, and Research Council of Norway (via grant no. 180439/V30 and DAADppp mobility project no. 199854)

N. Jain, B. Stiller, I. Khan, Ch. Marquardt, and G. Leuchs are with the Max Planck Institute for the Science of Light, Günther-Scharowsky-Str. 1/Bau 24, 91058 Erlangen, Germany and Friedrich-Alexander Universität Erlangen-Nuremberg (FAU), Staudtstrasse 7/B2, 91058 Erlangen, Germany. (e-mail: nitin.jain@mpl.mpg.de; birgit.stiller@mpl.mpg.de; imran.khan@mpl.mpg.de; christoph.marquardt@mpl.mpg.de; gerd.leuchs@mpl.mpg.de).

V. Makarov is with the Institute for Quantum Computing, University of Waterloo, Ontario N2L 3G1, Canada (e-mail: makarov@vad1.com).

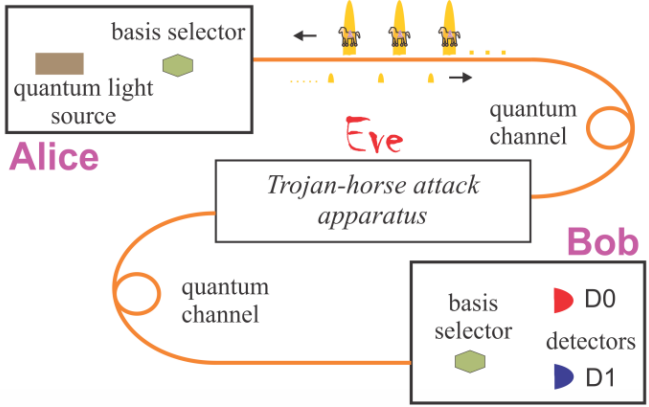

Fig. 1. Scheme of a Trojan-horse attack. Eve attacks Alice by sending bright Trojan-horse pulses to know the bases selected by the latter during the operation of the QKD protocol. This information is carried by the back-reflected pulses coming out of Alice. As a rule of thumb, Eve must avoid disturbing the legitimate quantum signals travelling from Alice to Bob as much as possible since she is interested in knowing only the basis settings. Only the most relevant components are shown in Alice and Bob for the sake of clarity. Fig. 6(a) shows a more detailed schematic of Bob.

for performing the attack in all these works is (in the vicinity of) 1550 nm, the standard telecom wavelength which is also employed by Alice and Bob for their communication. The feasibility and constraints of launching Trojan-horse attacks at different wavelengths have not been explored, especially from an experimental perspective. Such a study is crucial for a comprehensive analysis of the threat of these attacks and (re)design of suitable prevention mechanisms and countermeasures.

In this paper, we analyze the spectral behaviour of a variety of optical devices relevant to QKD, ranging from some single passive components to a complete QKD subsystem. In all these investigations, we also carefully assess the associated security risks from Trojan-horse attacks. The paper is organized as follows: in section II, we discuss a few concepts and ideas that define the Trojan-horse attack and also form the basis of its success or failure. In section III, we present the results of our experimental investigation into the spectral behaviour of two frequently-used components in practical QKD systems: a circulator and an isolator. Note that the latter is actually even a well-known countermeasure against Trojan-horse attacks on one-way QKD systems [23, 24]. We then describe some simple spectral measurements that were performed on the receiver system in Clavis2 (Bob), in the quest of improving the attack presented in Ref. [21]. In particular, we also show the results of our estimation of the spectral sensitivity of single-photon avalanche diodes (SPADs) in the 1700 - 1800 nm range. These SPADs are used by Bob to detect the quantum signals from Alice. In section IV and V, we sum up our work, provide an outlook, and finally discuss some countermeasures.

## II. BASIC CONCEPTS AND IDEAS

As illustrated in Fig.1, back-reflections that carry information of the applied bases are measured by Eve, for instance using state discrimination methods [29], as and when they return to the attack apparatus. If Eve could indeed do this entire operation without alerting Alice and Bob, she could break the security of any prepare-and-measure protocol (including both discrete and continuous variable protocols) [1, 30-32, 35]. However, as mentioned before, such vulnerability has been known [23, 24] for some time.

Two common practical measures to prevent or catch Trojan-horse attacks in action are to add an isolator or install a 'watchdog' detector, respectively, at the entrance of the Alice subsystem. While an ideal isolator would passively torpedo any Trojan-horse attack by a complete extinction of Eve's pulses – no matter how bright – dispatched into Alice, an ideal watchdog or monitoring detector would actively raise an alarm whenever any unknown or non-designated optical signals arrive into Alice.

However, while practical isolators have only a finite isolation, which may even be lower at wavelengths that are outside the 'design' wavelength bands, practical monitoring detectors have a vanishing responsivity outside a finite spectral range. In other words, there may exist a wavelength, or a set of wavelengths, at which an isolator can only partially stop a Trojan-horse pulse, or the response of the monitoring system is not sufficient to trigger the alarm. Additionally, if the back-reflection levels from one or more optical interfaces inside the QKD system at such wavelengths are also high, the risk of a successful Trojan-horse attack is naturally escalated. In the subsections below, we elaborate this risk further using some simple examples.

### A. Back-reflections and Eve's photon budget

In a Trojan-horse attack, Eve's light goes back and forth through the attacked subsystem. Moreover, at least one of the onward Trojan-horse pulse or the back-reflected pulse must probe or pass through the basis selector, e.g., a phase modulator [21]. The onward and reverse paths (which may not be the same necessarily [19]) decide the insertion loss, and together with the back-reflection level, determine the total attenuation suffered by Eve's pulse in the double pass. With the knowledge of these values, Eve can estimate the number of photons to expect on average ($\mu_{Eve}$), in the back-reflected pulse of interest as it travels to her on the quantum channel.

In the following, we explain the concept of Eve's photon budget by means of a numerical example. We assume the attack wavelength to be $\lambda \sim 1550$ nm, a binary basis choice, and that $\mu_{Eve} \approx 4$ suffices for accurately knowing the probed choice of the basis. We also assume the source of the back-reflection of interest to be an open fiber-optic interface, e.g., the glass-air interface of an unused port of a fiber coupler. This has actually been used for attacking a continuous variable QKD system [19]. For standard optical fiber components, a Fresnel reflection of around 4% (reflectivity $\mathcal{R} \approx -14$ dB) is obtained from such an interface. Let us assume that this coupler is located inside Alice (of a generic QKD system) in such a manner that the total insertion loss from the quantum channel to this coupler and back is $IL \approx -46$ dB. Given this, the total attenuation suffered by Eve's light can be calculated as $\mathcal{R} + IL \approx -60$ dB. Eve's Trojan-horse pulse needs to have roughly $4\times10^6$ photons (when it enters the QKD system) so that the back-reflected pulse of interest would carry the necessary 4 photons on average.

However, if Alice also uses an isolator providing an isolation (transmission in reverse direction) of ≥50 dB at λ ~ 1550 nm to protect the system against Trojan-horse attacks, Eve's photon budget needs to be raised by at least 5 orders of magnitude (>$4\times10^{11}$ photons). Such powerful pulses, apart from being quite easily detectable, may even be on the verge of the damage thresholds of the components or interconnects inside Alice. While a carefully-induced damage [7] may help the eavesdropper, in general, it is simply going to render the QKD device or the communication channel useless and thus does not benefit Eve.

### B. Foiling Alice's safeguards

The values of reflectivity, insertion loss, and isolation mentioned above are typically specified only for, or around, the design wavelength of the components. At other wavelengths, they are likely to differ. If the isolation is much lower at a wavelength $\lambda_{Eve}$ (far from 1550 nm), then Eve need not significantly raise the power of the Trojan-horse pulses. If Eve is fortunate, a higher reflectivity and/or lower insertion loss may be obtained at $\lambda_{Eve}$ resulting in a lower attenuation of her light through Alice. In such a case, the isolator would effectively be providing only a *false* sense of security because the peak power of Eve's Trojan-horse pulses would not be much higher than the value without the isolator (while $\mu_{Eve} \approx 4$ is still obtained).

As mentioned before, using a monitoring detector is another common method to catch real-time Trojan-horse attacks. In fact, they are indispensable for plug-and-play schemes since the light travels back and forth between Bob and Alice (implying that isolators cannot be used). For instance, a major part of the incoming light is diverted (using a 90/10 coupler) to an array of classical detectors in Clavis2-Alice [33]. Flaws which may allow the operation of these detectors to be manipulated by Eve have been recently discovered [20].

However, even without the flaws, a monitoring detector could be spectrally unresponsive outside a finite wavelength band. This increases the vulnerability against Trojan-horse attacks. To elaborate using another simple and fictitious example, let us assume that the monitoring detector has a negligible spectral sensitivity at $\lambda_{Eve}$, in particular, the responsivity $\rho(\lambda_{Eve}) < 10^{-4}$ A/W. Also, assume the minimum dark noise current $I_D$ of the detector is ~10 nA. If Eve sends a Trojan-horse pulse with a peak power[1] $P_{Eve}$ ~ 100 μW, the maximum resultant photocurrent in the classical detector is $I_{Ph} \leq \rho(\lambda_{Eve}) \times P_{Eve}$ = 10 nA. This implies the monitoring system's response to the Trojan-horse pulse is likely to get buried in the dark noise, allowing Eve to stay concealed.

The only other indicators of Eve's attack on a QKD system are an increased quantum bit error rate (QBER) and irregular detection statistics, and both of these quantities are typically estimated by Bob in classical postprocessing [2, 3]. A Trojan-horse attack on Alice – exploiting either inadequate isolation or insufficient monitoring system's response – is unlikely to affect Bob (and therefore the QBER and/or the detection statistics). In other words, Eve has a good chance to obtain information about the *complete* raw key without Alice's and Bob's knowledge.

### C. Avoiding discovery by Bob

Trojan-horse attacks have traditionally been known as a threat to only the Alice subsystem, as our discussion so far has also indicated. This is mainly due to the fact that in BB84, the most popular and widely-used QKD protocol, Bob anyway reveals his basis choice during the step of reconciliation [1-3]. Probing Bob's basis selection by means of Trojan-horse attack is therefore completely useless for Eve.

However, in the SARG04 protocol [34, 35], the secret bit is given by Bob's basis choice. As demonstrated experimentally in Ref. [21], this basis choice can be inferred with more than 90% accuracy even with a few back-reflected photons ($\mu_{Eve} \approx 3$). Nevertheless, the total attenuation suffered by Eve's pulse is of the order of −57 dB due to which, Eve has to send Trojan-horse pulses containing at least $1.5\times10^6$ photons into Bob. Unfortunately for Eve, Bob also contains a pair of gated SPADs and her bright pulses lead to a tremendous amount of afterpulsing [16] in these SPADs. Afterpulsing is caused due to the filling of the carrier traps which decay exponentially, causing uncorrelated clicks in a later gate [36, 37]. This elevates the dark noise level and hence, the QBER measured by Bob and Alice. As mentioned in the beginning, QKD systems normally abort the protocol if the QBER crosses a certain security threshold. Thus, it is strongly in Eve's interest to keep in check the increase in the afterpulsing probabilities in the SPADs due to her Trojan-horse pulses.

In some sense therefore, the SPADs in Bob act like a monitor against Trojan-horse attacks. Moreover, just like for the monitoring detector, the quantum efficiency of an SPAD is also a function of the wavelength of light impinging upon them. Typically, SPADs are designed to have their peak efficiency of detection in the telecom wavelength bands [38]. Although the physics that explains the concept of efficiency differs from that of afterpulsing, a lowered efficiency at a wavelength *longer* than the peak wavelength (i.e., λ > 1550 nm) should also imply a lower probability of afterpulsing. This conjecture is based on the fact that the longer-wavelength photons are less energetic to cross the bandgap (optimized for 1550 nm). The propensity to populate carrier traps and cause afterpulses should likewise be lesser. In fact, CW illumination around 1950 nm has even been used to depopulate the carrier traps by means of photoionization [39].

Eve could therefore choose a wavelength > 1550 nm to prepare the attack pulses (say $\lambda_{Eve}$ ~ 1700 nm), where the afterpulsing induced by the bright Trojan-horse pulses in the SPADs is considerably lower than at 1550 nm. She could even further mitigate the afterpulsing by adding CW light at λ ~ 1950 nm. An attack strategy exploiting other known features of the SPADs, e.g., deadtime, inability to resolve photon numbers, etc. [27] could then be used to enhance Eve's gain. To elaborate, Eve may obtain a significant portion of the raw key at a much reduced attack rate (that directly translates into a lowered QBER). In such cases, the privacy amplification performed by Alice and Bob at the end of the protocol may not suffice to erase Eve's knowledge. Such a result would have a major significance since SPADs are currently the most popular devices for single-photon detection in discrete variable QKD. Even more, entanglement-based protocol (including BB84) implementations can also be compromised.

---

[1] Depending on the optical pulse width T, the average number of photons per pulse is given by the expression $\lambda_{Eve} P_{Eve} T / \hbar c$

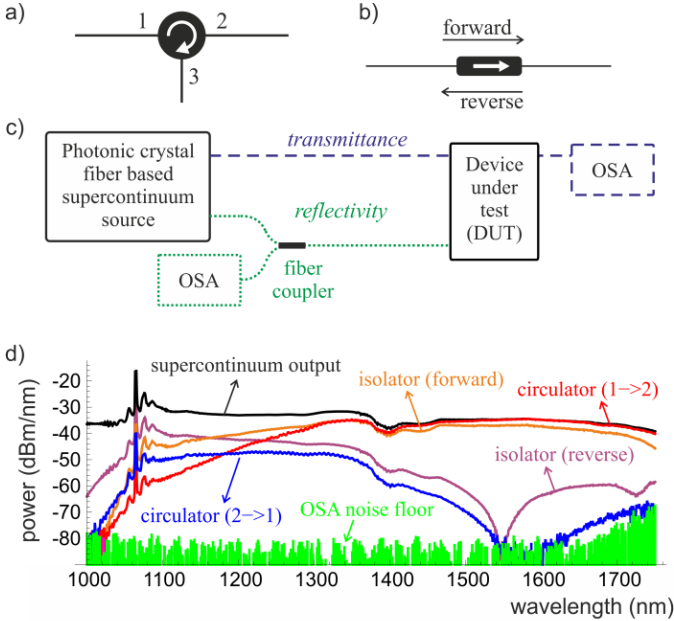

Fig. 2. Spectral transmission and measurement of light through different fiber-optical devices. a) The circulator allows optical transmission from port 1 to 2 and from port 2 to 3. Ideally, transmission is highly suppressed for the reverse directions. b) The isolator is supposed to have minimal attenuation in forward direction and maximum attenuation in reverse direction. c) We tested different devices (DUTs), from a single isolator to a complete QKD subsystem, with the objective of measuring the spectral transmittance or reflectivity of these devices. This was done by connecting them to an optical spectrum analyzer (OSA) and supercontinuum source in different configurations. d) The following spectral traces are shown: direct supercontinuum output (black); transmitted outputs of an isolator in both forward and reverse directions (orange and purple), output of the circulator at port 2 while the supercontinuum source is connected to port 1 and vice versa (red and blue). OSA model used for measurement: Ando AQ-6315A.

To illustrate the last points, let us consider a hypothetical example of a state-of-the-art QKD link where Alice and Bob share a secret key by measuring polarization entangled pairs as per the BB84 protocol. Let us assume that without Eve's attack, Bob and Alice incur a QBER $q_0 = 0.01$ and estimate the fraction of single-pair and uncorrelated multi-pair events [3] as $y_0 = 0.70$. As per security proofs [40-42], the QBER threshold above which Alice and Bob should abort the communication is $q_{abort} \approx 0.11$. Furthermore, assume that a variation $\delta y_{max} = 0.15$ in the estimation of y is allowed by the QKD system[2]. If the Trojan-horse attack, using the ideas discussed above, results in Eve obtaining correlations above 48% with the error-corrected key, and Alice and Bob observing the parameters under attack as $q_1 = 0.05$ and $y_1 =$ 0.79, then the security of the QKD system is violated. This is because $q_1 < q_{abort}$ and $\delta y$ (= $|y_1/y_0 - 1|$ = 0.13) < $\delta y_{max}$ would not raise any alarm, while the quantity to be subtracted during privacy amplification (calculated based on equation 40 in Ref. [3]) is 47.8% which is slightly smaller than Eve's knowledge.

## III. SPECTRAL MEASUREMENTS

In the previous section, we have developed the motivation to investigate possible security loopholes that arise from the behaviour of the QKD system at uncharacterized wavelengths and aid the Trojan-horse attack of Eve. In this section, we present our results on comprehensively testing the broadband transmittance and reflectivity of different fiber-optical components, namely a circulator and an isolator. We also perform some simple measurements on a whole receiver system (Clavis2-Bob). Finally, we also look into the spectral efficiency of SPADs in the 1700 - 1800 nm range, where the afterpulsing is conjectured to be low.

Fig. 2(a) and 2(b) show the schematic of a circulator and an isolator. Fig. 2(c) shows a schematic of the possible setups employed for the different measurements. For instance, to perform a spectral reflectivity measurement, the optical spectrum analyzer (OSA) could be connected to the device under test (DUT) via a balanced fiber splitter (green/dotted line). For a transmission measurement, the broadband light simply passes through the DUT (blue/dashed line).

As broadband source, we employed supercontinuum generation in a photonic crystal fiber [43]. A solid core photonic crystal fiber was pumped at a wavelength of 1064 nm to produce a broad and bright optical spectrum – the supercontinuum – stretching from below 600 nm to above 1700 nm [44]. The black trace in Fig. 2(d) shows the output in the 1000 - 1750 nm range, which we found to be the most interesting and relevant for a majority of the experiments presented in this section. The small dip around 1400 nm is typical for glass fibers due to infrared absorption. Nonetheless, the supercontinuum is > 40 dB above the noise floor of the optical spectrum analyzer (OSA) in the 1000 - 1650 nm range.

### A. Spectral transmittance of relevant fiber-optic components

We have measured the spectral characteristics of one fiber-optic circulator and three different fiber-optic isolators which are optimized for 1550 nm. As shown in Fig. 2(a), the circulator allows an optimal transmission from port 1 to port 2 and from port 2 to port 3 which is used, for instance, to separate counter-propagating signals. At 1550 nm, the minimum isolation, defined as the transmission from port 2 to port 1 or from port 3 to port2, as per the datasheet is 40 dB. The return loss is >60 dB and the insertion loss <1.1 dB.

Isolators were already introduced in the previous section and as illustrated in Fig. 2(b), they are normally used as a one-way optical transmission component. As per the datasheets of three tested isolators, it should be possible to obtain a minimum isolation (transmission in reverse direction) of 40 dB, return loss of >55 dB, and insertion loss of <1.0 dB.

---
[2] This value of $\delta y_{max}$ is used by Clavis2 for example.

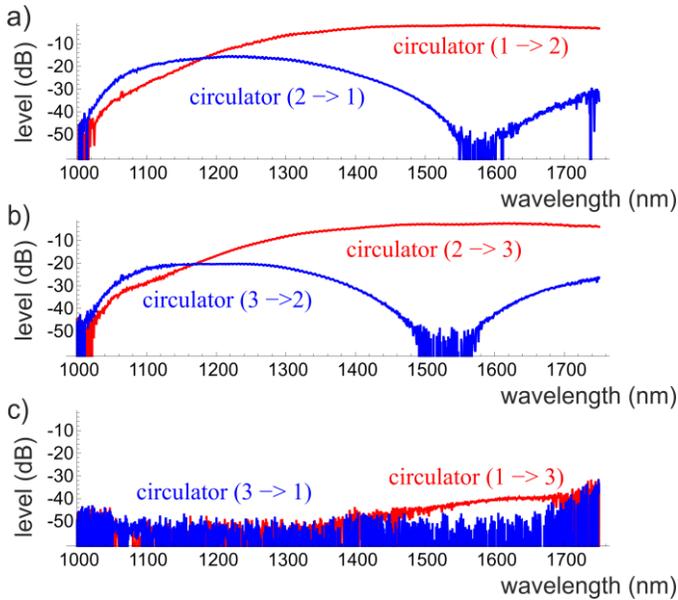

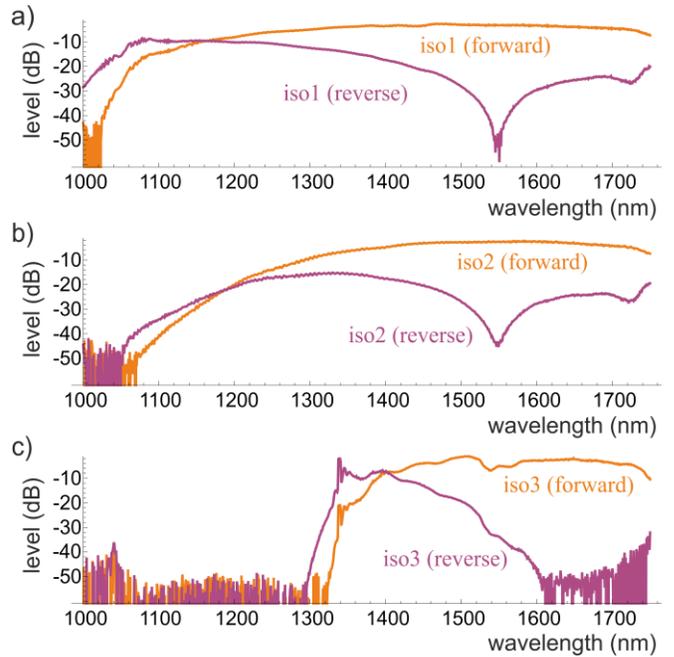

Fig. 3. Inferred transmittance of an optical circulator at different ports. The raw traces (not shown here) were obtained using the bright supercontinuum output trace shown in Fig. 2(d). The spectral traces shown here are normalized to the output of the supercontinuum source.

Fig. 4. Inferred transmittance of the three isolators in the forward and reverse directions. The raw traces (not shown here) were obtained in the same way as in Fig. 3 and the traces shown here are also normalized to the output of the supercontinuum source.

The broadband source was connected to the circulator and isolator, the devices under tests (DUTs) one by one; see the dashed/blue line in Fig. 2(c). All possible directions of transmission through these DUTs were investigated. Fig. 2(d) shows some absolute spectral traces that can be readily compared with the supercontinuum output. The raw output spectra/traces for the forward and reverse transmission for port 1 and 2 of the circulator are shown in red and blue, respectively. Similarly, the raw traces for the transmission through forward and reverse directions of an isolator are respectively given by the yellow and purple traces.

As expected, the reverse direction of the circulator (2→1) exhibits high attenuation in comparison to the favoured direction (1→2), especially between 1550 nm and 1600 nm. Nevertheless, we can observe a promising behaviour, for example, around 1310 nm, where the transmission through the reverse direction is significantly higher (~40 dB) compared to that at 1550 nm. Similarly, the extinction ratio between the forward and reverse direction of the isolator is significant at 1550 nm, roughly −50 dB, but around 1300 nm, the power transmitted in the reverse direction is merely 10 dB lower than that in the forward direction.

For a more accurate comparison, we normalized the various transmission spectra to the supercontinuum source output. The resulting relative spectral traces are shown in Fig. 3 for the circulator and in Fig. 4 for the three isolators. In Fig. 3(a) and 3(b), the favoured directions from port 1 to 2, and from port from port 2 to 3 show a similar transmittance. The insertion losses are around −2 dB for port 1 to 2, and −2.8 dB for port 2 to 3. The maximum power transfer is observed in the 1400 - 1750 nm regime. This can be exploited to receive a back-reflection of interest via the reverse direction of a circulator. The spectral transmission of light through port 1 to 3 and vice-versa shows high attenuation all over the spectrum. There is a slight transmission from port 1 to port 3 in the 1450 - 1700 nm range, but the transmittance is about −40 dB; combined with other insertion losses, this is unlikely to aid Eve.

Fig. 4 presents the normalized transmission traces of the three isolators. We do not have details about the fabrication process but we can distinguish them by our measurements, as is evident in Fig. 4. In the forward direction, at wavelengths longer than 1400 nm, the transmittance of all three isolators matches the specified insertion loss values (of around 0.5 dB). In the reverse direction, the attenuation levels range from −50 dB (iso1) to −35 dB (iso3) at 1550 nm. Also, for iso1 and iso2, the maximum attenuation is observed only in a rather narrow dip around 1550 nm. But the attenuation level at 1310 nm in the reverse direction for iso1 is at least 30 dB lower than that at 1550 nm. This is certainly promising for Trojan-horse attacks. For iso3, the traces in both the forward and reverse directions are markedly different from their counterparts of iso1 and iso2. In general, another attack wavelength would be more appropriate for iso3 since the observed transmittance in both forward and reverse directions is quite low at 1310 nm.

In Alice, an isolator is sometimes used just after the laser source in order to prevent back-reflections into the laser. Such a configuration could also be interesting for Eve if the isolator shows high back-reflections for some wavelengths. Therefore, we measured the reflectivity of the aforementioned isolators. Unfortunately (for Eve), we did not find any useful reflection signal which would have aided the Trojan-horse attack.

As the principle of Trojan-horse attacks is based on sending a bright pulse and analyzing the back-reflected signal, the quantity ultimately relevant to deciding Eve's photon budget see section II.A) is the net transmittance in a double pass through the fiber-optic components. Fig. 5 shows this net transmittance of a pulse seen in a double pass, which means that the pulse has passed twice through the DUT, first in the forward and then in the reverse direction or vice versa. For the circulator (Fig. 5(a)), we observe a maximum attenuation at the telecom wavelength, as expected. However, the minimum attenuation suffered in a double pass around 1300 nm is about −25 dB, which seems reasonable for an attack. Note that the pulse also suffers a higher loss in the forward direction when using a wavelength other than 1550 nm.

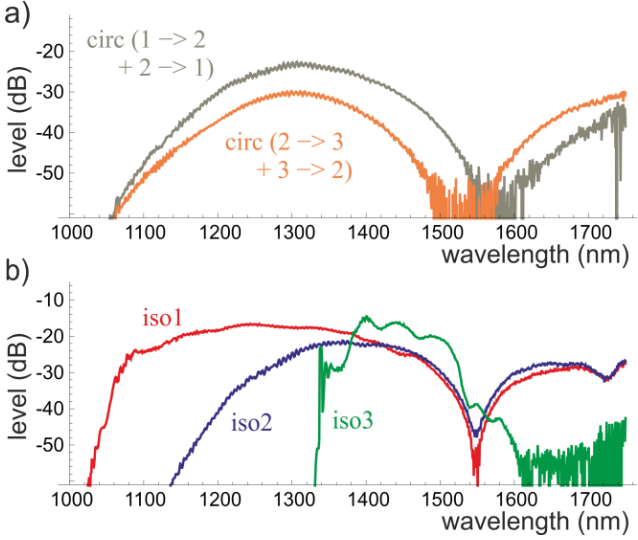

Fig. 5. Net transmittance observed in a double pass a) through the circulator that was characterized in Fig. 3, and b) through the three isolators that were characterized in Fig. 4.

In Fig. 5(b), the spectral traces for the isolators reveal large differences in terms of the net transmittance in a double pass. While iso3 provides the lowest attenuation at around 1400 nm (−15 dB), its isolation at 1550 nm also shows the worst performance amongst the three isolators. The other two isolators show a broader spectral transmittance: for iso1, it interestingly is transmitting all the way to 1100 nm. The maximal isolation at 1550 nm is −45 dB (iso2) and −50 dB (iso1), but this is only in a narrow regime around 1550 nm. It is thus clear that even high performance isolators do not have high isolation in other wavelength regions, such as from 1300 to 1400 nm, where Eve can easily obtain both laser sources and detectors for performing an efficient Trojan-horse attack. This result is however not very surprising since most isolators rely on the wavelength-dependent Faraday effect for their correct functionality.

### B. Reflection & transmission measurements on Clavis2-Bob

In section II.C, we explained some of the problems and constraints encountered in attacking Clavis2-Bob at 1550 nm. In short, Bob contains the single-photon avalanche diodes (SPADs) that incur a strong afterpulsing from the Trojan-horse pulses, which reveals the presence of Eve. This is due to the total attenuation suffered by Eve in going back and forth through Bob being fairly high (around −57 dB at 1550 nm).

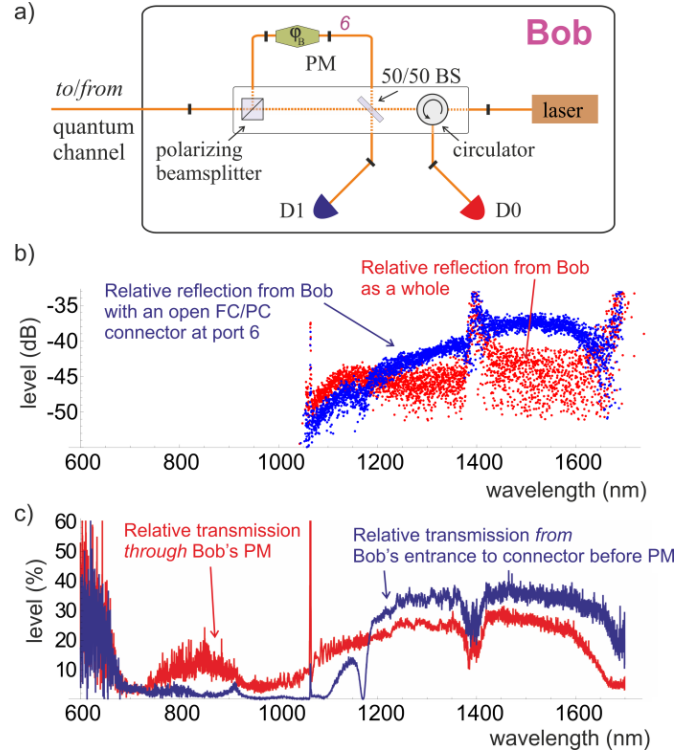

Fig. 6. Results of simple transmission and reflection measurements on Clavis2-Bob. (a) A supercontinuum source, with an output similar to the one shown in Fig. 2(c), was used. a) Simplified optical schematic of Bob. Various optic/optoelectronic components are connected with fibers (solid orange lines) using FC/PC connectors, denoted here by the small black rectangular blocks. The polarizing beamsplitter, 50/50 BS (beamsplitter), and circulator form a composite assembly in which light propagates in free space (dotted orange line). b) The sum total of the reflected light coming out of Bob (red trace) shows a perceptible variation when the FC/PC connector (port 6 in Fig. 6(a)) after the phase modulator (PM) is opened; see the blue trace. The traces are relative, i.e. they have been normalized to the original supercontinuum output. At around 1550 nm, the ~8 dB rise in the back-reflection (blue trace, compared to red trace) can be attributed to a strong Fresnel reflection from the connector. c) Relative transmission measurements mainly highlight the effects of absorption until and inside Bob's PM. All traces were measured by an optical spectrum analyzer (model: Anritsu AQ6317B).

Eve sends less than $2\times10^6$ photons per pulse into Bob, yet, the subsequent afterpulsing results in a huge dark count rate.

However, as also explained in section II, both reflectivity of the different components and afterpulsing probability in an SPAD may depend on Eve's wavelength. To elaborate, if Eve could send dimmer Trojan-horse pulses at a wavelength where she still obtains a decent back-reflection of interest (i.e., $\mu_{Eve}$ still remains high enough to yield reasonable success during Eve's measurement of the back-reflection; see section II.A.) and where the afterpulsing is additionally lower, the chances of being caught by Bob are reduced.

In Ref. [21], we performed optical time domain reflectometry (OTDR) measurements for three different wavelengths: 808 nm, 1310 nm, and 1550 nm. Ideally, one should repeat such OTDR measurements at a variety of wavelengths to check for the reflectivity dependence. However, such a procedure would be extremely cumbersome and time-consuming. To keep matters simple, we decided to test the whole subsystem of Bob using a broadband source and an optical spectrum analyzer (OSA). In other words, Clavis2-Bob was our device under test (DUT). Fig. 6(a) shows a simplified optical schematic of Clavis2-Bob. The aim of the Trojan-horse attack *per se* is to know $\phi_B$, the phase modulated by Bob on the quantum signals from Alice.

We re-employed a supercontinuum source, similar to the one used earlier (see Fig. 2), for this purpose. For the reflection measurements, the supercontinuum was inserted into Bob via a pre-characterized coupler and the reflection traces were obtained by plugging the other input port of the coupler into the OSA; see the green/dotted line in Fig. 2(c). The absolute back-reflection measurement traces (not shown here) were obtained for two cases: with Bob as a whole, and with the output FC/PC connector of Bob's phase modulator (PM) additionally disconnected. Fig. 6(b) shows the traces normalized to the original supercontinuum just like in Fig. 3 and 4. The traces until 1064 nm are intentionally blanked out since the corresponding reflections were buried in noise floor of the OSA. Fig. 6(c) shows the transmittance also for two cases: from Bob's entrance to the input FC/PC connector of Bob's PM and through Bob's PM. For all these measurements, the input polarization was adjusted to transmit the maximum amount of light into Bob's PM.

A sharp peak (dip) in the spectral reflectivity (transmittance) at some wavelength in the scanned range could have been of a great interest to an eavesdropper for launching a Trojan-horse attack. In that regard, the transmittance in Fig. 6(c) does show a drop around 1170 nm (see the blue curve). However, there is no corresponding peak in the reflectivity traces. It is likely therefore that this dip comes from absorption (at the polarizing beamsplitter cube) instead of reflection.

*C. Spectral efficiency of a single-photon avalanche diode*

In section II.C, we conjectured about the lowering of the afterpulsing probability in gated single-photon avalanche diodes (SPADs) at wavelengths longer than 1550 nm. Briefly, an SPAD optimized to detect photons at 1550 nm would display a cut-off wavelength $\lambda_C > 1550$ nm because a photon at $\lambda$ beyond $\lambda_C$ would not have sufficient energy to cross the bandgap [45, 46]. A lower spectral sensitivity may be seen as a potential indicator of lower afterpulsing: if the SPAD response at a wavelength far from 1550 nm is characterized by much smaller detection probabilities *inside* the gate, then the SPAD might prove to be less prone to afterpulses created *outside* the gate [16, 21].

Based on this premise, we tried to measure the sensitivity of the SPADs in Clavis2-Bob in the 1700 - 1800 nm range, i.e., at wavelengths likely to be above the cut-off wavelength. We used the pulsed idler beam output from an optical parametric oscillator (OPO) for this purpose. The OPO could be scanned and configured to emit pulses at some desired wavelength in the 1680 - 2000 nm range.

The OPO is not, or rather, cannot be synchronized to the SPAD gates. Also, it operates at a repetition rate of ~ 82 MHz which is much higher than the gating frequency (set to ~ 98.0 kHz for this specific experiment [12]) of the SPADs in Bob. One may thus assume that the SPAD sees the light from the OPO as quasi-CW essentially. Based on this assumption, one can calculate the mean photon number $\mu = \tau \lambda P_i / \hbar c$ seen inside a gate. Here $\tau \approx 2.5$ ns denotes the full width at half maximum value of the gate applied on the SPAD, $\lambda$ and $P_i$ are the wavelength and average power of the idler output, respectively. With the setup connected to one of the detectors in Bob, we recorded the number of clicks $n_c$ and number of gates $N_g$ to evaluate the total detection probability:

$$p^{tot} = \frac{n_c}{N_g} = d + p_\mu^{ph}(1-d)$$

where $d$ is the dark count probability (known value), and $p_\mu^{ph} = 1 - e^{-\mu \eta(\lambda)}$ is the photonic detection probability. The spectral sensitivity $\eta(\lambda)$ of the SPAD can therefore be obtained by inverting the above equation.

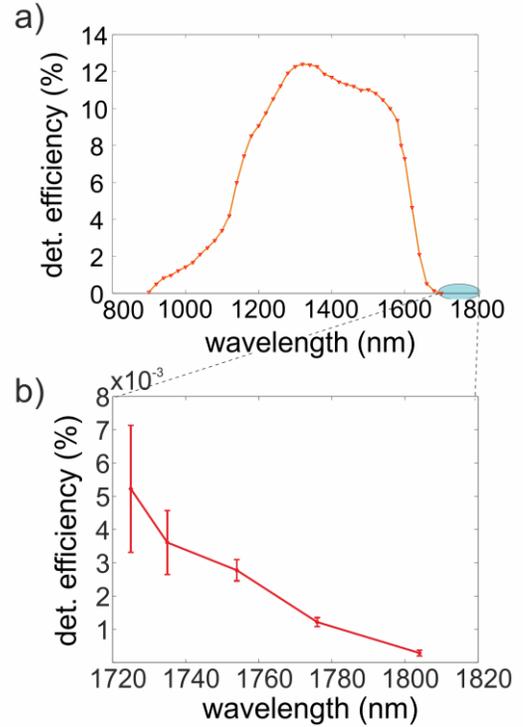

Fig. 7. Spectral sensitivity of the SPAD. a) Spectral data of a typical SPAD from ID Quantique [17]. The shaded ellipse shape is essentially zoomed in b) which is based on our measurement results with the idler beam. The sensitivity of the SPAD in the 1700 - 1800 nm range is at least three orders of magnitude lower than the peak value around 1310 nm.

Fig. 7(b) shows the results of our measurements and calculations (both SPADs had an almost similar spectral response, so we show the result for only one of them). We recorded the total detection probabilities at each wavelength shown for different values of the mean photon number $\mu$ by variably attenuating the idler beam.

Compared to the response at 1550 nm, where $\eta \approx 0.1$ is a typical value as shown in Fig. 7(a), the SPADs are indeed much less spectrally sensitive in the tested regime. At $\lambda \approx$ 1720 nm, the efficiency $\eta \approx 0.05 \times 10^{-3}$ indicates that it would be worth trying to measure the actual afterpulsing response at this wavelength. Although it must be noted that insertion losses through fiber-optic components at $\lambda \sim 1720$ nm are higher than at 1550 nm. To maintain the photon budget, as explained in section II.A, more energetic pulses would be required which would increase the afterpulsing probability. This tradeoff needs to be studied more thoroughly.

## IV. SUMMARY AND OUTLOOK

To summarize, we have discussed how the wavelength dependence of optical devices – frequently used in practical fiber-optical QKD systems – can make the system more vulnerable to Trojan-horse attacks. In that regard, we have pointed out the practical limits of two measures specifically advertised for circumventing Trojan-horse attacks. These countermeasures are passive or active devices, such as an isolator or monitoring detector at Alice's entrance, that can prevent or catch Trojan-horse pulses as and when these pulses try to enter into the QKD subsystem.

In particular, we have measured the spectral transmittance of different isolators. While the performance in and around the design wavelength of 1550 nm is pretty much as expected, an increased risk of Trojan-horse attacks is seen at wavelengths far from 1550 nm. For example, Eve gets a benefit of almost 30 dB if she chooses her attack wavelength to be around 1300 nm. Insertion losses of most telecom components in this wavelength regime also do not differ much from that around 1550 nm. In fact, in a round trip of Eve's light, the total attenuation at wavelengths around 1300 nm or 1700 nm can be considerably lesser than at 1550 nm, as attested by the spectral analysis of circulators. We also note that performing a spectral characterization of the QKD system from the quantum channel is a one-time job for Eve that can be performed rather quickly, i.e., at the most, it may distract Alice/Bob only momentarily.

To obtain a live signature of such attacks, Alice may also use a monitoring detector. Although we could not perform a spectral measurement of the behaviour of a classical monitoring detector, we did measure the spectral sensitivity of single-photon avalanche diodes (SPAD). The context here was similar: a straightforward Trojan-horse attack on Bob fails due to the severe afterpulsing caused by the bright pulses from Eve. We conjectured that a lower sensitivity at a wavelength *longer* than the peak wavelength (i.e., $\lambda > 1550$ nm) should also imply a lower probability of afterpulsing. Our experimental measurements indicate that the sensitivity of an SPAD in Bob in the 1700 - 1800 nm range is at least three orders of magnitude lower than the peak value. Although fiber attenuation is higher around 1700 nm, typical QKD receivers do not contain long fiber links. Also, as our measurement results have shown, the behaviour of some components such as isolators and circulators do not seem too inimical for Eve.

Based on these findings, the security of a QKD system that uses *only* an isolator (in the transmitter, Alice) for protection against Trojan-horse attacks could be breached successfully by sending bright pulses around 1300 nm. Likewise, an attack on the receiver Bob, who contains single-photon detectors, should have better chances of remaining undetected if Eve prepares bright pulses around 1700 nm. Note that the attack against isolators at 1310 nm is applicable to both discrete and continuous variable QKD systems. We believe trying out proof-of-principle Trojan-horse attacks on practical QKD systems [19, 21] around these wavelengths (1300 nm or 1700 nm) is the best way to check the effectiveness of existing and new countermeasures, and to reduce the threat of these attacks convincingly.

## V. COUNTERMEASURES

Both general and specific countermeasures against Trojan-horse attacks have been discussed before [19-21, 23, 24]. Here we discuss some measures for countering Trojan-horse attacks based on the discussion in the previous sections.

If Alice contains a monitoring detector in *addition* to the isolator, then it would become fairly challenging for Eve to simultaneously circumvent both of these countermeasures. To make it even more difficult, Alice could additionally insert an optical filter based on fiber Bragg gratings (FBGs) or Fabry-Perot cavities, after the monitoring detector. Filters based on fiber Bragg gratings for example can be designed with different filtering widths and central wavelengths, and depending on the filter design [47], operate in the reflection mode. The transmission spectrum of commercially-available filters reported in datasheets is typically characterized in a range of 20 nm around the central wavelength. Within this range, the FBG transmission profile might contain secondary maxima which must be taken into account. However, out of this wavelength range, where there is a higher risk of an attack as our measurements have shown, the filters feature a high suppression of the spectral components.

With such an optical filter, if a pulse from Eve at some wavelength $\lambda_{Eve}$ is neither made extinct by the isolator nor caught by the monitoring detector, then it would still not reach the basis selector as it would be filtered out before. A filtering width of a few nm should suffice to ensure undisturbed operation of the QKD system since all sidebands of the phase and intensity modulation would be within the bandwidth of the filter. Such a filter would also prove beneficial for the security of Bob.

From a more general perspective, methods that reduce the intensity of back-reflections – occurring inside either Alice or Bob – also naturally contribute to the overall security of the system. Some practical examples are: using angle polished connectors (FC/APC) instead of flat connectors (FC/PC), re-designing the overall system to eliminate any open ports [19], and fusing all connections if and when possible (instead of using mating sleeves). Note that these steps can limit the chances of an attack independent of the spectral response of the single optical devices.

Finally, note that in most instances of Trojan-horse attacks, Eve gains only a partial but non-negligible amount of the secret key. From a theoretical perspective, a higher amount of privacy amplification can always help Alice and Bob to destroy the partial information of Eve. Nonetheless, to quantify the requisite amount of privacy amplification, one must carefully scrutinize each subsystem, estimate the maximum leakage due to Trojan-horse attacks, and incorporate these elements in the security proof.

## VI. Acknowledgements

We would like to thank Alessio Stefani, Mohiudeen Azhar, Nicolas Joly, and Philip St. J. Russell for their help with the supercontinuum sources.

## References

[1] C. H. Bennett and G. Brassard, "Quantum Cryptography: Public Key Distribution and Coin Tossing" in *Proc. IEEE International Conference on Computers Systems and Signal Processing*, Bangalore, India, 1984, pp. 175

[2] N. Gisin, G. Ribordy, W. Tittel and H. Zbinden. (2002, Mar.). Quantum cryptography. *Reviews of Modern Physics* [Online]. *74(1)*, pp. 145–195. Available: http://journals.aps.org/rmp/abstract/10.1103/RevModPhys.74.145

[3] V. Scarani, H. Bechmann-Pasquinucci, N. Cerf, M. Dušek, N. Lütkenhaus and M. Peev. (2009, Sep.). The security of practical quantum key distribution. *Reviews of Modern Physics* [Online]. *81(3)*, pp. 1301–1350. Available: http://journals.aps.org/rmp/abstract/10.1103/RevModPhys.81.1301

[4] N. Jain, "Security of practical quantum key distribution systems", Ph.D. dissertation (in progress), Max Planck Institute for the Science of Light, Erlangen, Germany, 2014

[5] V. Makarov, "Quantum cryptography and quantum cryptanalysis", Ph.D. dissertation, Norwegian University of Science and Technology, Trondheim, Norway, 2007

[6] L. Lydersen, "Practical security of quantum cryptography", Ph.D. dissertation, Norwegian University of Science and Technology, Trondheim, Norway, 2011

[7] N. Bugge, S. Sauge, A. M. M. Ghazali, J. Skaar, L. Lydersen and V. Makarov. (2014, Feb.). Laser Damage Helps the Eavesdropper in Quantum Cryptography. *Physical Review Letters* [Online]. *112(7)*, pp. 070503. Available: http://journals.aps.org/prl/abstract/10.1103/PhysRevLett.112.070503

[8] I. Gerhardt, Q. Liu, A. Lamas-Linares, J. Skaar, C. Kurtsiefer and V. Makarov. (2011, Jan.). Full-field implementation of a perfect eavesdropper on a quantum cryptography system. *Nature Communications* [Online]. *2(2027)*, pp. 349. Available: http://www.nature.com/ncomms/journal/v2/n6/full/ncomms1348.html

[9] N. Jain, C. Wittmann, L. Lydersen, C. Wiechers, D. Elser, Ch. Marquardt, V. Makarov and G. Leuchs. (2011, Sep.). Device Calibration Impacts Security of Quantum Key Distribution. *Physical Review Letters* [Online]. *107(11)*, pp. 110501. Available: http://journals.aps.org/prl/abstract/10.1103/PhysRevLett.107.110501

[10] A. Lamas-Linares and C. Kurtsiefer. (2007, Apr.). Breaking a quantum key distribution system through a timing side channel. *Optics Express* [Online]. *15(15)*, pp. 9388. Available: http://www.opticsinfobase.org/oe/abstract.cfm?uri=oe-15-15-9388

[11] H.-W. Li, S. Wang, J.-Z. Huang, W. Chen, Z.-Q. Yin, F.-Y. Li, Z. Zhou, D. Liu, Y. Zhang, G.-C. Gou, W.-S. Bao and Z.-F. Han. (2011, Dec.). Attacking a practical quantum-key-distribution system with wavelength-dependent beam-splitter and multiwavelength sources. *Physical Review A* [Online]. *84(6)*, pp. 062308. Available: http://journals.aps.org/pra/abstract/10.1103/PhysRevA.84.062308

[12] L. Lydersen, N. Jain, C. Wittmann, Ø. Marøy, J. Skaar, Ch. Marquardt, V. Makarov and G. Leuchs. (2011, Sep.). Superlinear threshold detectors in quantum cryptography. *Physical Review A* [Online]. *84(3)*, pp. 032320. Available: http://journals.aps.org/pra/abstract/10.1103/PhysRevA.84.032320

[13] L. Lydersen, C. Wiechers, C. Wittmann, D. Elser, J. Skaar and V. Makarov. (2010, Aug.). Hacking commercial quantum cryptography systems by tailored bright illumination. *Nature Photonics* [Online]. *4(10)*, pp. 686–689. Available: http://www.nature.com/nphoton/journal/v4/n10/full/nphoton.2010.214.html

[14] S. Nauerth, M. Fürst, T. Schmitt-Manderbach, H. Weier and H. Weinfurter. (2009, Jun.). Information leakage via side channels in freespace BB84 quantum cryptography. *New Journal of Physics* [Online]. *11(6)*, pp. 065001. Available: http://iopscience.iop.org/1367-2630/11/6/065001/

[15] S.-H. Sun, M.-S. Jiang and L.-M. Liang. (2011, Jun.). Passive Faraday-mirror attack in a practical two-way quantum-key-distribution system. *Physical Review A* [Online]. *83(6)*, pp. 062331. Available:

[16] C. Wiechers, L. Lydersen, C. Wittmann, D. Elser, J. Skaar, Ch. Marquardt, V. Makarov and G. Leuchs. (2011, Jan.). After-gate attack on a quantum cryptosystem. *New Journal of Physics* [Online]. *13(1)*, pp. 013043. Available: http://iopscience.iop.org/1367-2630/13/1/013043/

[17] F. Xu, B. Qi and H.-K. Lo. (2010, Nov.). Experimental demonstration of phase-remapping attack in a practical quantum key distribution system. *New Journal of Physics* [Online]. *12(11)*, pp. 113026. Available: http://iopscience.iop.org/1367-2630/12/11/113026/

[18] Y. Zhao, C.-H. Fung, B. Qi, C. Chen and H.-K. Lo. (2008, Oct.). Quantum hacking: Experimental demonstration of time-shift attack against practical quantum-key-distribution systems. *Physical Review A* [Online]. *78(4)*, pp. 042333. Available: http://journals.aps.org/pra/abstract/10.1103/PhysRevA.78.042333

[19] I. Khan, N. Jain, B. Stiller, P. Jouguet, S. Kunz-Jacques, E. Diamanti, Ch. Marquardt and G. Leuchs, "Trojan-horse attacks on practical continuous-variable quantum key distribution systems", QCRYPT, Paris, Sep. 2014

[20] S. Sajeed, I. Radchenko, S. Kaiser, J.-P. Bourgoin, L. Monat, M. Legré and V. Makarov, "Securing two-way quantum communication: the monitoring detector and its flaws", QCRYPT, Paris, Sep. 2014

[21] N. Jain, E. Anisimova, I. Khan, V. Makarov, Ch. Marquardt and G. Leuchs. "Trojan-horse attacks threaten the security of practical quantum cryptography", arXiv: 1406.5813 (accepted for publication in New Journal of Physics)

[22] D. S. Bethune and W. P. Risk. (2000, Mar.). An autocompensating fiber-optic quantum cryptography system based on polarization splitting of light. *IEEE Journal of Quantum Electronics* [Online]. *36(3)*, pp. 340–347. Available: http://ieeexplore.ieee.org/xpl/articleDetails.jsp?arnumber=825881

[23] A. Vakhitov, V. Makarov and D. R. Hjelme. (2001, Nov.). Large pulse attack as a method of conventional optical eavesdropping in quantum cryptography. *Journal of Modern Optics* [Online]. *48(13)*, pp. 2023–2038. Available: http://www.tandfonline.com/doi/abs/10.1080/09500340108240904

[24] N. Gisin, S. Fasel, B. Kraus, H. Zbinden and G. Ribordy. (2006, Feb.). Trojan-horse attacks on quantum-key-distribution systems. *Physical Review A* [Online]. *73(2)*, pp. 022320. Available: http://journals.aps.org/pra/abstract/10.1103/PhysRevA.73.022320

[25] M. Bourennane, F. Gibson, A. Karlsson, A. Hening, P. Jonsson, T. Tsegaye, D. Ljunggren and E. Sundberg. (1999, May). Experiments on long wavelength (1550 nm) "plug and play" quantum cryptography systems. *Optics Express* [Online]. *4(10)*, pp. 383. Available: http://www.opticsinfobase.org/oe/abstract.cfm?uri=oe-4-10-383

[26] D. Stucki, N. Gisin, O. Guinnard, G. Ribordy and H. Zbinden. (2002, Jul.). Quantum key distribution over 67 km with a plug&play system. *New Journal of Physics* [Online]. *4(1)*, pp. 41. Available: http://iopscience.iop.org/1367-2630/4/1/341/

[27] ID Quantique, http://www.idquantique.com/

[28] SeQureNet, http://www.sequrenet.com/

[29] C. W. Helstrom. (1969, Mar.). Quantum detection and estimation theory. *Journal of Statistical Physics* [Online]. *1(2)*, pp 231. Available: http://link.springer.com/article/10.1007%2FBF01007479

[30] F. Grosshans and P. Grangier. (2002, Jan.). Continuous Variable Quantum Cryptography Using Coherent States. *Physical Review Letters* [Online]. *88(5)*, pp. 057902. Available: http://link.aps.org/doi/10.1103/PhysRevLett.88.057902

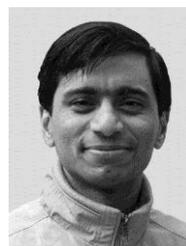

**Nitin Jain** was born in Gurgaon, India in 1981. He obtained B.Tech in electrical engineering and M.Tech in communications & signal processing from IIT Bombay (India) in 2004. He worked in the software industry for a while, developing video codecs for handheld devices at NVIDIA Graphics Pvt. Ltd., Pune (India). In 2007, he headed to the University of Calgary (Canada) to do M.Sc. in physics with main focus on quantum information science and technology.

Since 2010, he has been pursuing a doctorate at the Max Planck Institute for the Science of Light in Erlangen (Germany). His thesis work is on the security of practical quantum key distribution systems. His research interests span both fundamental and applied aspects of quantum information apart from sundry topics such as optical detection, code optimization, and cryptography.

Mr. Jain was the recipient of the National Talent Search Exam (NTSE) award in 1997 and the best poster award at the WE-Heraeus conference in 2010. In 2006, he was recognized amongst the top 10 percent of performers in NVIDIA worldwide. He is currently a member of German Physical Society (DPG) and SPIE.

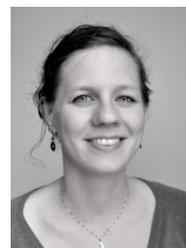

**Birgit Stiller** received her M.Sc. degree in Physics and Mathematics from the Friedrich-Alexander University of Erlangen-Nuremberg, Germany, in 2008, as a Cusanuswerk scholarship recipient. In 2011 she obtained her Ph.D. degree in Nonlinear Fiber Optics at the CNRS Research Institute FEMTO-ST, Besançon, France, where she continued as a postdoctoral researcher in 2012.

Since October 2012 she is a postdoctoral fellow at the Max Planck Institute for the Science of Light in Erlangen, Germany. Her research interests include nonlinear fiber optics, in particular Brillouin scattering and phase-sensitive amplifiers, nonlinear effects in photonic crystal fibers, as well as quantum communication, specifically quantum receivers and secure communication.

Dr. Stiller was awarded the Ohm Prize for her Master thesis in 2009 and the A'Doc Prize for her doctoral thesis in 2011. She is a member of the German Physical Society (DPG).

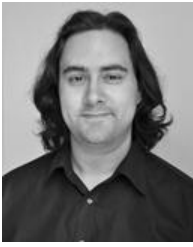
**Imran Khan** was born in Nürnberg, Germany, in 1986. He received his diploma in physics from the University of Erlangen-Nuremberg, Erlangen, in 2011.

Since 2011, he has been a PhD candidate at the Max Planck Institute for the Science of Light in Erlangen. His research interests include quantum optics, specifically quantum communication with applications in secure communication, random number generation and its underlying technologies.

Mr. Khan is a member of the German Physical Society (DPG) and part of the OSA student chapter Erlangen.

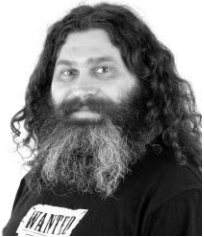
**Vadim Makarov** was born in Leningrad, USSR in 1974. He received the B.S. and M.S. degrees in radiophysics from St. Petersburg State Polytechnical University, Russia in 1998 and the Ph.D. degree in quantum cryptography from the Norwegian University of Science and Technology, Trondheim, Norway in 2007.

He was a postdoc at the Pohang University of Science and Technology, Pohang, South Korea in 2007--2008, then at the Norwegian University of Science and Technology in 2008--2011. Since 2012, he has been a research assistant professor leading the Quantum hacking lab at the Institute for Quantum Computing, University of Waterloo, Canada. His research interests are centered around practical security of quantum cryptography systems, and technology of free-space and satellite-based quantum communications.

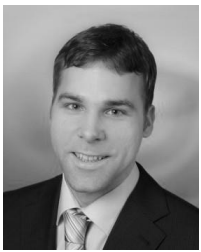
**Christoph Marquardt** was born in Berlin, Germany in 1976. He received his Dipl.-Phys. and Dr. rer. nat. degree at the University of Erlangen-Nuremberg in 2002 and 2007. In 2008, he worked as a metrology scientist at Carl Zeiss Laser Optics GmbH and then returned to the University of Erlangen-Nuremberg. From June 2012 to April 2014, he held a two year position as Alcatel Lucent Bell Labs guest professor. Currently he is a permanent staff at the Max Planck Institute for the Science of Light, where he is the leader of the quantum information processing group.

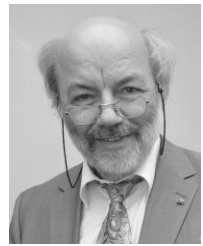
**Gerd Leuchs** studied Physics and Mathematics at the University of Cologne, Germany, and received his PhD in 1978. After two research visits at the University of Colorado in Boulder, USA, he headed the German gravitational wave detection group from 1985 to 1989. He then became the technical director at Nanotech AG in Switzerland. Since 1994, Professor Leuchs has been holding the chair for optics at the Friedrich-Alexander-University of Erlangen-Nuremberg, Germany. His fields of research span from aspects of classical optics to quantum optics and quantum information.

Since 2003, he has been the director of the Max Planck Research Group for Optics, Information and Photonics at Erlangen, which became the newly-founded Max Planck Institute for the Science of Light at Erlangen in 2009. He has served as an editor of 3 books, authored more than 300 articles published in peer reviewed scientific journals and 10 patented inventions. Since 2012 he is also an adjunct Professor at the University of Ottawa in Canada.